\documentclass[twocolumn,prl]{revtex4}

\usepackage{amsfonts}
\usepackage{graphics}
\usepackage{epsfig}
\usepackage{amscd}

\begin{document}

\title{String-nets, single and double-stranded quantum loop gases for non-Abelian anyons}
\author{Andrea Velenich}
\affiliation{Physics Department, Boston University, Boston, MA 02215, U.S.A.}
\author{Claudio Chamon}
\affiliation{Physics Department, Boston University, Boston, MA 02215, U.S.A.}
\author{Xiao-Gang Wen}
\affiliation{Department of Physics, Massachusetts Institute of Technology, Cambridge, MA 02215, U.S.A.}
\date{\today}

\begin{abstract}

String-net condensation can give rise to non-Abelian anyons whereas loop condensation usually gives rise to Abelian anyons.  It has been proposed that generalized quantum loop gases with non-orthogonal inner products can produce non-Abelian anyons.  We detail an exact mapping between the string-net and the generalized loop models and explain how the non-orthogonal products arise.
We also introduce a loop model of double-stranded nets where quantum loops with an orthogonal inner product and local interactions supports non-Abelian Fibonacci anyons. Finally we emphasize the origin of the sign problem in such systems and its consequences on the complexity of their ground state wave functions.

\end{abstract}

\maketitle

Some strongly correlated quantum many body systems display a type of order which is topological in nature and cannot be characterized by local order parameters~\cite{Wen}. Fractional Quantum Hall systems are the primary examples so far. A robust ground state degeneracy which cannot be lifted by any local perturbation~\cite{Haldane1985,Wen1990} and fractionalized degrees of freedom~\cite{Laughlin,Halperin,Arovas} are among the exotic properties which might be exploited for decoherence-free topological quantum computation~\cite{K03,NSSFD08}.

Topological phases with Abelian quasiparticles in two spatial dimensions can be constructed from loop models whose ground states are condensates of fluctuating non-intersecting closed loops. Non-Abelian phases are more difficult to attain, one route being to consider string-nets, which do allow for intersections of strings~\cite{LW04}. Alternatively, Fendley has recently proposed that non-Abelian phases could also be obtained from quantum loop gases without intersections if one allows for non-orthogonal inner products between classical loop configurations \cite{F08,TTSN08}. In this paper we demonstrate explicitly that such a quantum loop gas formulation follows directly from the string-net rules, establishing precisely the connection between the two approaches. We also introduce an example of a \emph{loop} model with an \emph{orthogonal} inner product supporting non-Abelian excitations. Finally we show that attempts to endow the loops with a dynamics {\it \`a la} Rokhsar-Kivelson, using simple moves between pairs of loop configuration, are doomed to fail.
In fact, the Hamiltonians are plagued by a ``sign problem'' which hinders a simple study of non-Abelian topological states using, for instance, standard Monte Carlo methods.

\emph{String-net rules as projectors}. The ground state wave function of a string-net model for Fibonacci anyons \cite{FF05,FFNWW,TTWL08} must satisfy the following constraints \cite{LW04}:
\begin{eqnarray} \label{XGrules3}
&\Phi \Big( \begin{picture}(18,-50) \put(1,-6){\includegraphics[height=0.6cm]{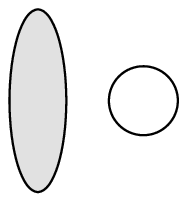}} \end{picture} \Big) = \gamma \Phi \Big( \begin{picture}(9,-50) \put(1,-6){\includegraphics[height=0.6cm]{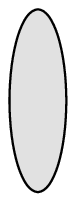}} \end{picture} \Big) \\
\label{XGrules}
& \Phi \Big( \begin{picture}(27,-50) \put(1,-6){\includegraphics[height=0.6cm]{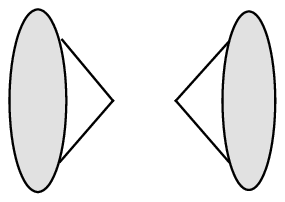}} \end{picture} \Big) = \gamma^{-1} \Phi \Big( \begin{picture}(22,-50) \put(1,-6){\includegraphics[height=0.6cm]{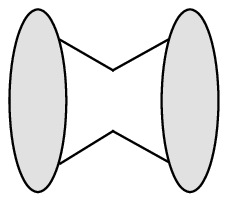}} \end{picture} \Big) + \gamma^{-\frac{1}{2}} \Phi \Big( \begin{picture}(22,-50) \put(1,-6){\includegraphics[height=0.6cm]{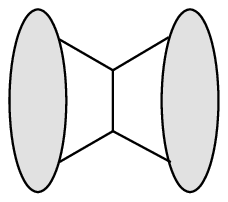}} \end{picture} \Big) \\
& \Phi \Big( \begin{picture}(27,-50) \put(1,-6){\includegraphics[height=0.6cm]{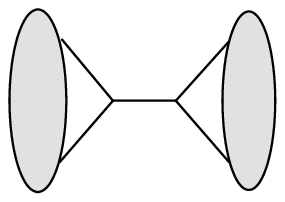}} \end{picture} \Big) = \gamma^{-\frac{1}{2}} \Phi \Big( \begin{picture}(22,-50) \put(1,-6){\includegraphics[height=0.6cm]{XG2.eps}} \end{picture} \Big) - \gamma^{-1} \Phi \Big( \begin{picture}(22,-50) \put(1,-6){\includegraphics[height=0.6cm]{XG4.eps}} \end{picture} \nonumber \Big)
\end{eqnarray}
where $\gamma = \frac{1+\sqrt{5}}{2}$ and $\gamma^2 = \gamma + 1$.
The ``surgery'' relations can also be inverted to yield:
\begin{eqnarray} \label{XGrules2}
& \Phi \Big( \begin{picture}(22,-50) \put(1,-6){\includegraphics[height=0.6cm]{XG2.eps}} \end{picture} \Big) = \gamma^{-1} \Phi \Big( \begin{picture}(27,-50) \put(1,-6){\includegraphics[height=0.6cm]{XG1.eps}} \end{picture} \Big) + \gamma^{-\frac{1}{2}} \Phi \Big( \begin{picture}(27,-50) \put(1,-6){\includegraphics[height=0.6cm]{XG3.eps}} \end{picture} \Big) \\
& \Phi \Big( \begin{picture}(22,-50) \put(1,-6){\includegraphics[height=0.6cm]{XG4.eps}} \end{picture} \Big) = \gamma^{-\frac{1}{2}} \Phi \Big( \begin{picture}(27,-50) \put(1,-6){\includegraphics[height=0.6cm]{XG1.eps}} \end{picture} \Big) - \gamma^{-1} \Phi \Big( \begin{picture}(27,-50) \put(1,-6){\includegraphics[height=0.6cm]{XG3.eps}} \end{picture} \nonumber \Big) 
\end{eqnarray}
After being rescaled by a factor $-1/2$, (\ref{XGrules}) and (\ref{XGrules2}) together are conveniently expressed in matrix form as:
\begin{equation} \label{defF}
F \, \mathbf{v} = \frac{1}{2} \left( \begin{array}{cccc}
1	& -\gamma^{-1}		& 0	& -\gamma^{-\frac{1}{2}}	\\
-\gamma^{-1}	& 1	& -\gamma^{-\frac{1}{2}}	& 0	\\
0	& -\gamma^{-\frac{1}{2}}	& 1	& \gamma^{-1}	\\
-\gamma^{-\frac{1}{2}}		& 0	& \gamma^{-1}	& 1
\end{array} \right) \mathbf{v} = 0
\end{equation}
where $\mathbf{v}
= v_1 | \begin{picture}(12,-40) \put(2,-2){\includegraphics[width=0.3cm]{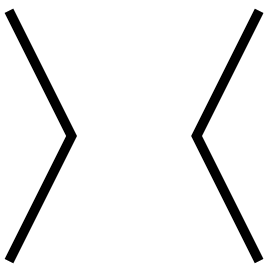}} \end{picture} \rangle 
+ v_2 | \begin{picture}(12,-40) \put(2,-2){\includegraphics[width=0.3cm]{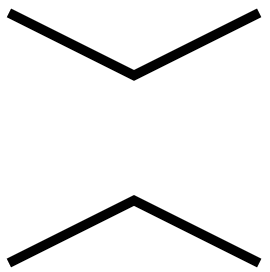}} \end{picture} \rangle 
+ v_3 | \begin{picture}(12,-40) \put(2,-2){\includegraphics[width=0.3cm]{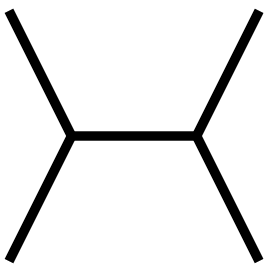}} \end{picture} \rangle 
+ v_4 | \begin{picture}(12,-40) \put(2,-2){\includegraphics[width=0.3cm]{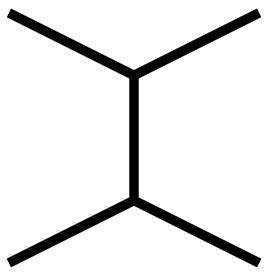}} \end{picture} \rangle$;
$v_1,v_2,v_3,v_4$ are the amplitudes 
$\Phi(\begin{picture}(16,7) \put(0,-3){\includegraphics[height=0.4cm]{XG1.eps}} \end{picture}), \Phi(\begin{picture}(13,7) \put(0,-3){\includegraphics[height=0.4cm]{XG2.eps}} \end{picture}), \Phi(\begin{picture}(16,7) \put(0,-3){\includegraphics[height=0.4cm]{XG3.eps}} \end{picture}), \Phi(\begin{picture}(13,7) \put(0,-3){\includegraphics[height=0.4cm]{XG4.eps}} \end{picture})$ respectively
and the vectors $\{ 
| \begin{picture}(12,-40) \put(2,-2){\includegraphics[width=0.3cm]{conf1.eps}} \end{picture} \rangle,
| \begin{picture}(12,-40) \put(2,-2){\includegraphics[width=0.3cm]{conf2.eps}} \end{picture} \rangle,
| \begin{picture}(12,-40) \put(2,-2){\includegraphics[width=0.3cm]{conf3.eps}} \end{picture} \rangle,
| \begin{picture}(12,-40) \put(2,-2){\includegraphics[width=0.3cm]{conf4.eps}} \end{picture} \rangle \}$, 
labelled by classical string-net configurations, are assumed to be an \emph{orthonormal} basis of the Hilbert space.
$F$ is Hermitian and $F^2 = F$, hence $F$ is a projector; the eigenvalues are $\{0,0,1,1 \}$ with corresponding eigenvectors $\{| g_1 \rangle, | g_2 \rangle, | e_3 \rangle, | e_4 \rangle \}$.
By construction, the 2-dimensional ground state spanned by $|g_1\rangle$ and $|g_2\rangle$ is a quantum superposition of classical string-net states satisfying the rules in (\ref{XGrules}) or (\ref{XGrules2}), whereas the excited states $|e_3\rangle$ and $|e_4\rangle$ violate such rules.
The local 4-dimensional projector in (\ref{defF}) is the minimal operator annihilating both $|g_1\rangle$ and
$|g_2\rangle$ and the ground state of the Fibonacci string-net model must be annihilated by a sum of these projectors implementing all the possible local surgeries.

\emph{Quantum loop models and non-orthogonal basis vectors.} The 4-dimensional formalism described above can be reduced in a natural way to a 2-dimensional one describing the appropriate Hilbert space for quantum loop models \cite{F08}.
Exploiting the identity $\frac{\gamma}{\sqrt{\gamma+1}}=1$ we can multiply the off-diagonal elements of the matrix $F$ in (\ref{defF}) by $\frac{\gamma}{\sqrt{\gamma+1}}$ and re-write it as:
\begin{equation} \label{Falpha}
F_{\alpha=\gamma} = \frac{1}{2} \left( \begin{array}{cccc}
1 & -\sqrt{\frac{1}{1+\gamma}} & 0 & -\sqrt{\frac{\gamma}{1+\gamma}} \\
-\sqrt{\frac{1}{1+\gamma}} & 1 & -\sqrt{\frac{\gamma}{1+\gamma}} & 0 \\
0 & -\sqrt{\frac{\gamma}{1+\gamma}} & 1 & \sqrt{\frac{1}{1+\gamma}} \\
-\sqrt{\frac{\gamma}{1+\gamma}} & 0 & \sqrt{\frac{1}{1+\gamma}} & 1
\end{array} \right) \nonumber
\end{equation}
Such a form is convenient for replacing the golden ratio $\gamma$ by an arbitrary \emph{positive} real number $\alpha$ (a choice of inner product for loop states will prove equivalent to fixing $\alpha$). $F_{\alpha}$ is still a projector with eigenvalues $\{ 0, 0, 1, 1 \}$ and corresponding ($\alpha$-dependent) orthonormal eigenvectors $\{ |g_1\rangle, |g_2\rangle, |e_3\rangle, |e_4\rangle \}$:
\begin{equation} \label{changebasis}
\left( \begin{array}{c}
g_1 \\
g_2 \\
e_3 \\
e_4
\end{array} \right) =
\left( \begin{array}{cccc}
a & a & b & b \\
-c & c & d & -d \\
-c & -c & d & d \\
a & -a & b & -b
\end{array} \right)
\left( \begin{array}{c}
\includegraphics[width=0.3cm]{conf1.eps} \\
\includegraphics[width=0.3cm]{conf2.eps} \\
\includegraphics[width=0.3cm]{conf3.eps} \\
\includegraphics[width=0.3cm]{conf4.eps}
\end{array} \right) \nonumber
\end{equation}
\begin{eqnarray}
& a = \frac{\sqrt{1+\alpha}+1}{2\sqrt{1+\alpha+\sqrt{1+\alpha}}} \quad ; \quad 
  b = \frac{\sqrt{\alpha}}{2\sqrt{1+\alpha+\sqrt{1+\alpha}}} \nonumber \\
& c = \frac{\sqrt{1+\alpha}-1}{2\sqrt{1+\alpha-\sqrt{1+\alpha}}} \quad ; \quad 
  d = \frac{\sqrt{\alpha}}{2\sqrt{1+\alpha-\sqrt{1+\alpha}}} \nonumber
\end{eqnarray}
The ground state of $F_{\alpha}$ is a 2-dimensional linear subspace of the original 4-dimensional space.
It is then natural to study the fate of the four orthonormal basis vectors $\{ 
| \begin{picture}(12,-40) \put(2,-2){\includegraphics[width=0.3cm]{conf1.eps}} \end{picture} \rangle,
| \begin{picture}(12,-40) \put(2,-2){\includegraphics[width=0.3cm]{conf2.eps}} \end{picture} \rangle,
| \begin{picture}(12,-40) \put(2,-2){\includegraphics[width=0.3cm]{conf3.eps}} \end{picture} \rangle,
| \begin{picture}(12,-40) \put(2,-2){\includegraphics[width=0.3cm]{conf4.eps}} \end{picture} \rangle \}$ 
when projected onto the 2-dimensional plane spanned by $|g_1\rangle$ and $|g_2\rangle$.
Such projected states will be called ``shadow'' states and marked by a hat.
Inverting (\ref{changebasis}) and rescaling by a factor of $\sqrt{2}$ to yield normalized shadow states, we obtain:
\begin{equation}
\begin{picture}(70,42) 
\put(2,-27){\includegraphics[width=1.5cm]{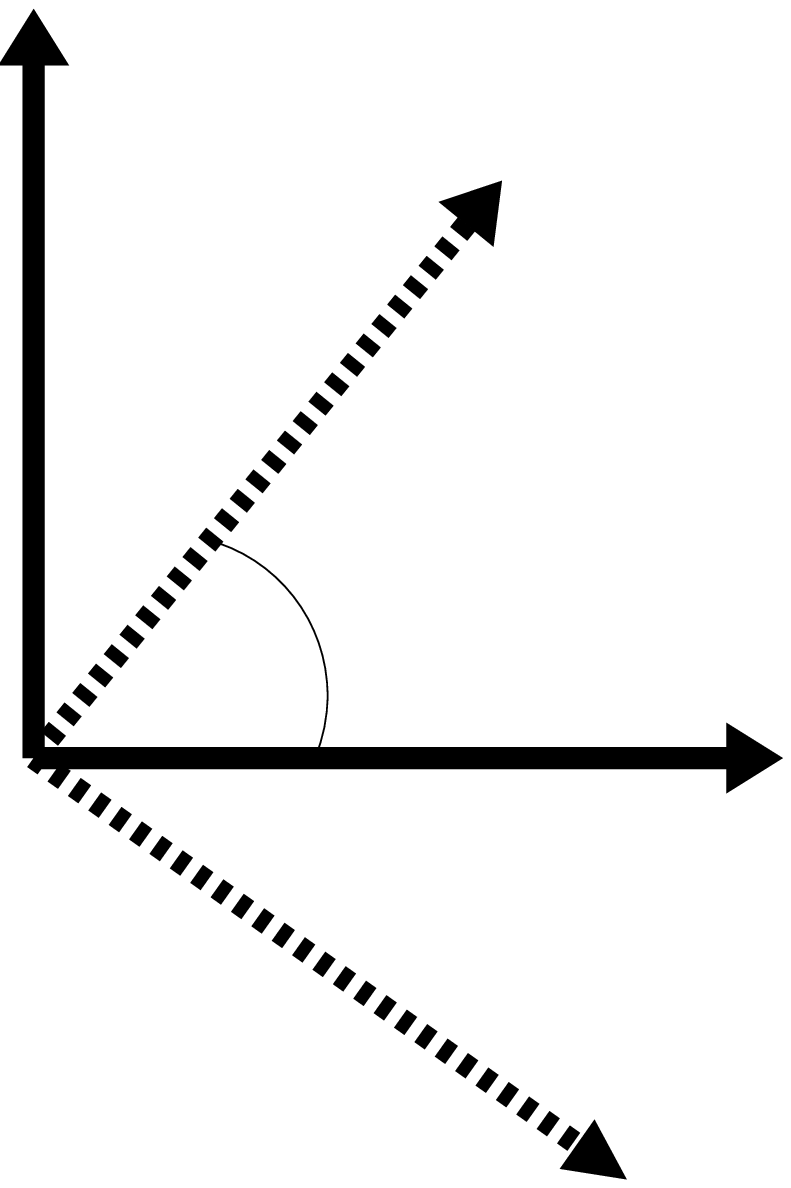}}
\put(45,-8){$\hat{\includegraphics[width=0.3cm]{conf1.eps}}$}
\put(32,23){$\hat{\includegraphics[width=0.3cm]{conf2.eps}}$}
\put(7,31){$\hat{\includegraphics[width=0.3cm]{conf3.eps}}$}
\put(36,-29){$\hat{\includegraphics[width=0.3cm]{conf4.eps}}$}
\end{picture} \quad \left( \begin{array}{c}
\hat{\includegraphics[width=0.3cm]{conf1.eps}} \\
\hat{\includegraphics[width=0.3cm]{conf2.eps}} \\
\hat{\includegraphics[width=0.3cm]{conf3.eps}} \\
\hat{\includegraphics[width=0.3cm]{conf4.eps}}
\end{array} \right) = \sqrt{2}
\left( \begin{array}{cc}
a & -c \\
a & c \\
b & d \\
b & -d
\end{array} \right)
\left( \begin{array}{c}
g_1 \\
g_2 \\
\end{array} \right)
\end{equation}
The inner product of the original 4-dimensional Hilbert space naturally induces an inner product on the 2-dimensional plane which, for every $\alpha \geq 0$, acts on the shadow vectors as:
\begin{eqnarray}
& \langle \hat{\begin{picture}(13,7) \put(2,-2){\includegraphics[width=0.3cm]{conf1.eps}} \end{picture}} 
| \hat{\begin{picture}(12,7) \put(2,-2){\includegraphics[width=0.3cm]{conf3.eps}} \end{picture}} \rangle 
= 2(ab-cd) = 0 \nonumber \\
& \langle \hat{\begin{picture}(13,7) \put(2,-2){\includegraphics[width=0.3cm]{conf2.eps}} \end{picture}} 
| \hat{\begin{picture}(12,7) \put(2,-2){\includegraphics[width=0.3cm]{conf4.eps}} \end{picture}} \rangle 
= 2(ab-cd) = 0 \nonumber \\
& \langle \hat{\begin{picture}(13,7) \put(2,-2){\includegraphics[width=0.3cm]{conf1.eps}} \end{picture}} 
| \hat{\begin{picture}(12,7) \put(2,-2){\includegraphics[width=0.3cm]{conf2.eps}} \end{picture}} \rangle = 2(a^2-c^2) = \frac{1}{\sqrt{1+\alpha}} \nonumber \\
& \langle \hat{\begin{picture}(13,7) \put(2,-2){\includegraphics[width=0.3cm]{conf3.eps}} \end{picture}} 
| \hat{\begin{picture}(12,7) \put(2,-2){\includegraphics[width=0.3cm]{conf4.eps}} \end{picture}} \rangle = 2(b^2-d^2) = -\frac{1}{\sqrt{1+\alpha}}  \nonumber
\end{eqnarray}
Only two shadow states are necessary for a basis of the 2-dimensional plane. One possibility is to choose one of the two orthogonal pairs $\{ 
| \hat{\begin{picture}(12,6) \put(2,-2){\includegraphics[width=0.3cm]{conf1.eps}} \end{picture}} \rangle,
| \hat{\begin{picture}(12,6) \put(2,-2){\includegraphics[width=0.3cm]{conf3.eps}} \end{picture}} \rangle \}$ and $\{
| \hat{\begin{picture}(12,6) \put(2,-2){\includegraphics[width=0.3cm]{conf2.eps}} \end{picture}} \rangle,
| \hat{\begin{picture}(12,6) \put(2,-2){\includegraphics[width=0.3cm]{conf4.eps}} \end{picture}} \rangle \}$. Alternatively, considering only states without branchings, the choice $\{
| \hat{\begin{picture}(12,6) \put(2,-2){\includegraphics[width=0.3cm]{conf1.eps}} \end{picture}} \rangle,
| \hat{\begin{picture}(12,6) \put(2,-2){\includegraphics[width=0.3cm]{conf2.eps}} \end{picture}} \rangle \}$ represents the natural basis of a quantum loop model. The substitution $\sqrt{1+\alpha}\leftrightarrow d=\lambda^{-1}$ ($d$ and $\lambda$ defined in \cite{F08}) provides the dictionary establishing the exact correspondence between the quantum loop states described in \cite{F08}, and the Fibonacci string-net states of \cite{LW04}. Hence the relations (\ref{XGrules}) and (\ref{XGrules2}) can be interpreted either as relations between \emph{amplitudes} in the 4-dimensional space where $\{ 
| \begin{picture}(12,-40) \put(2,-2){\includegraphics[width=0.3cm]{conf1.eps}} \end{picture} \rangle,
| \begin{picture}(12,-40) \put(2,-2){\includegraphics[width=0.3cm]{conf2.eps}} \end{picture} \rangle,
| \begin{picture}(12,-40) \put(2,-2){\includegraphics[width=0.3cm]{conf3.eps}} \end{picture} \rangle,
| \begin{picture}(12,-40) \put(2,-2){\includegraphics[width=0.3cm]{conf4.eps}} \end{picture} \rangle \}$
are orthonormal, or as relations between shadow \emph{basis vectors} in the 2-dimensional space where $\{ 
| \hat{\begin{picture}(12,6) \put(2,-2){\includegraphics[width=0.3cm]{conf1.eps}} \end{picture}} \rangle,
| \hat{\begin{picture}(12,6) \put(2,-2){\includegraphics[width=0.3cm]{conf2.eps}} \end{picture}} \rangle,
| \hat{\begin{picture}(12,6) \put(2,-2){\includegraphics[width=0.3cm]{conf3.eps}} \end{picture}} \rangle,
| \hat{\begin{picture}(12,6) \put(2,-2){\includegraphics[width=0.3cm]{conf4.eps}} \end{picture}} \rangle \}$
are not orthogonal.

\emph{Lattice implementation.}
Having proven the equivalence of string-nets and quantum loop gases in the continuum, we now discuss their \emph{microscopic} correspondence in a specific lattice system.
We focus on a simple fully-packed model on the square lattice, where four tile configurations are allowed \begin{picture}(12,10) \put(0,-3){\includegraphics[width=0.4cm]{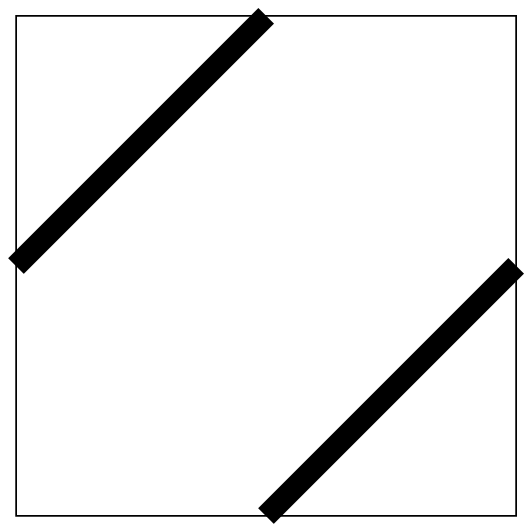}} \end{picture} \begin{picture}(12,10) \put(0,-3){\includegraphics[width=0.4cm]{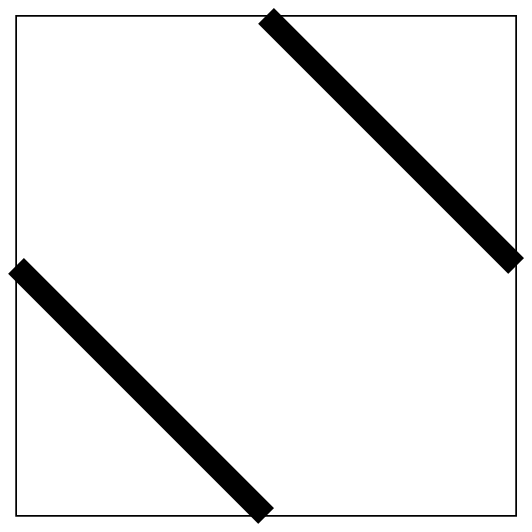}} \end{picture} \begin{picture}(12,10) \put(0,-3){\includegraphics[width=0.4cm]{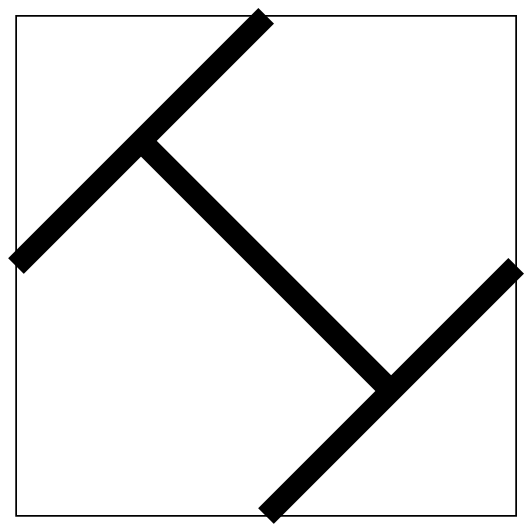}} \end{picture} \begin{picture}(12,10) \put(0,-3){\includegraphics[width=0.4cm]{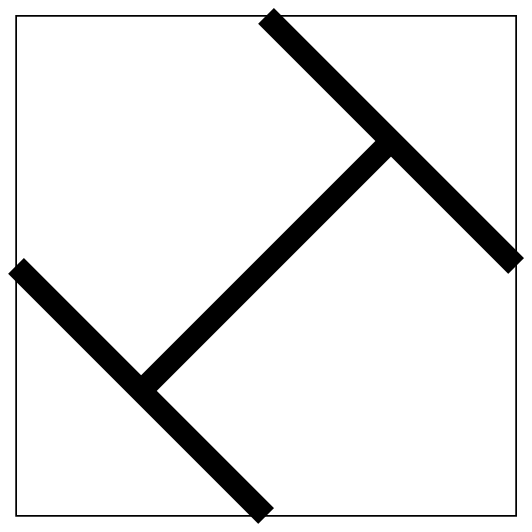}} \end{picture}, each tile representing a possible surgery point.
By introducing suitable dynamical rules we will show how this tile model can implement the Levin-Wen string-net model for Fibonacci anyons \cite{LW04}. We will then introduce two mappings: one into a non-orthogonal loop model {\it \`a la} Fendley and one into a loop model with an \emph{orthogonal} inner product.

The Hilbert space of an $N$-tile system is the N-fold tensor product of 4-dimensional local Hilbert spaces and the string-net rules can be applied at the single plaquette level so that the transition amplitudes between plaquette configurations are given by $F_{\alpha}$ in (\ref{Falpha}):
\begin{equation} \label{HF}
H_F = J \left( \mathbb{I} \otimes \ldots \otimes \mathbb{I} \otimes F_{\alpha} + \ldots + F_{\alpha} \otimes \mathbb{I} \otimes \ldots \otimes \mathbb{I} \right)
\end{equation}
By applying (\ref{HF}), the $4^N$-dimensional Hilbert space is projected onto the $2^N$-dimensional Hilbert space of $N$ decoupled spins 1/2, consistently with the fact that $F_{\alpha}$ leaves two effective degrees of freedom per tile. The redundancy of keeping four states per plaquette will be useful in the following to construct the mappings between sting-net and loop models.
Notice that enforcing the topological rules at the single tile level through (\ref{HF}) cannot be sufficient to drive the system into a topological phase since the full-packing condition in the model does not allow a string configuration on a single tile to stretch and spread its topology over more tiles.
The role of operators acting on several plaquettes at the same time, such as the the 12-spin operators in the following, is to enforce topological constraints on the degrees of freedom emerging from the fully-packed background so that the topological features of a configuration can propagate throughout the system. This turns out to be the key feature in arguing that \emph{local} lattice operators are sufficient to endow the system with topological properties at large scales.
FIG.1a through FIG.1c show how the Levin-Wen string-net model defined on the honeycomb lattice \cite{LW04} can be represented in our square tile language.
Down-spins (\includegraphics[width=0.2cm]{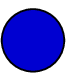}) represent string segments and up-spins (\includegraphics[width=0.2cm]{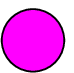}) represent empty edges.
The string-net configurations are constrained by 3-spin ``star'' operators (imposing the fusion rules) and the dynamics is implemented through 12-spin interactions (red dashed lines in FIG.1).
After deforming the honeycomb lattice to a brick wall lattice, we obtain a square lattice by adding edges populated by dummy up-spins (\includegraphics[width=0.2cm]{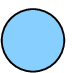}) (FIG.1b).
\begin{figure}[!h] \label{lattices}
\includegraphics[width=4.2cm]{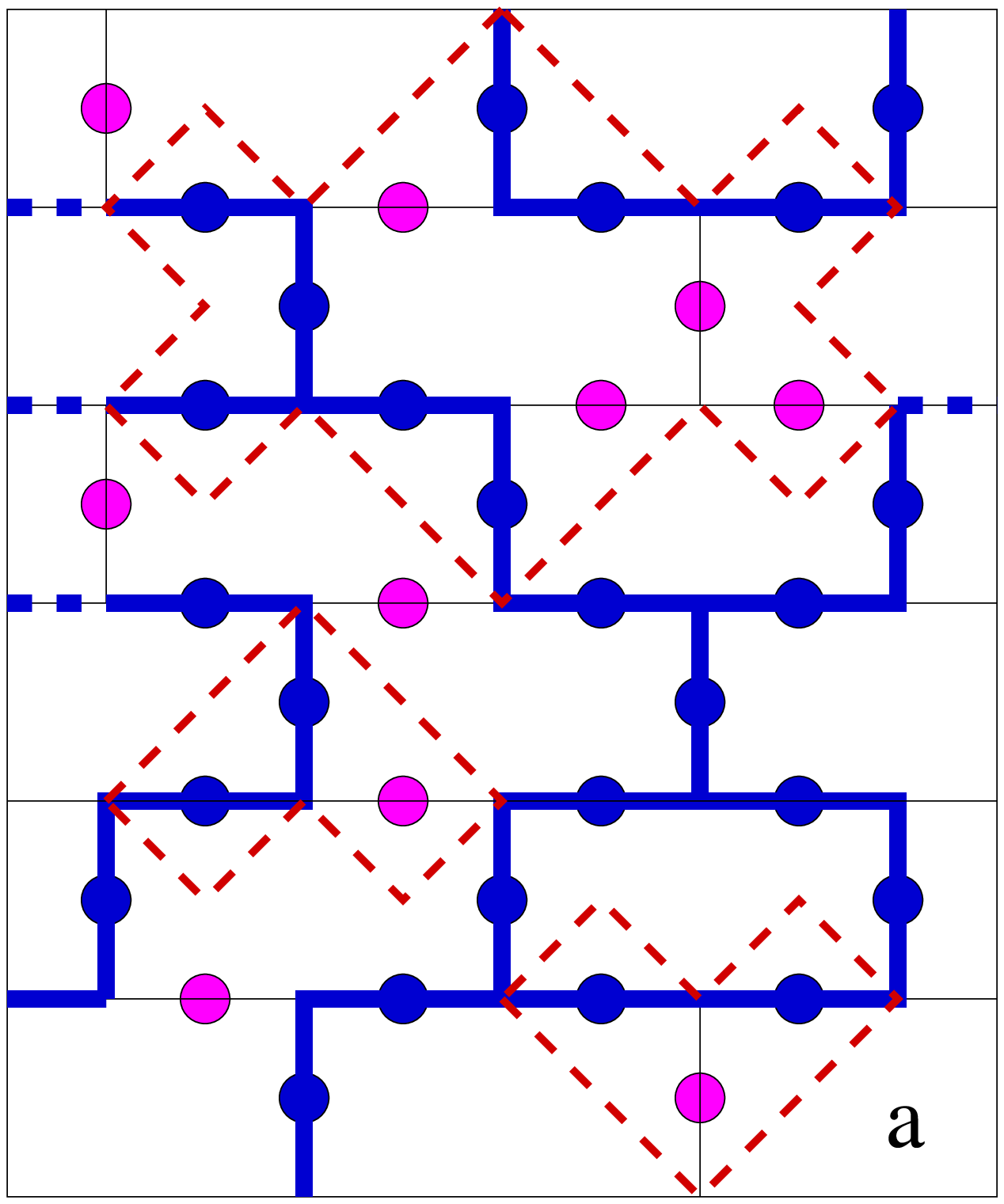} \includegraphics[width=4.2cm]{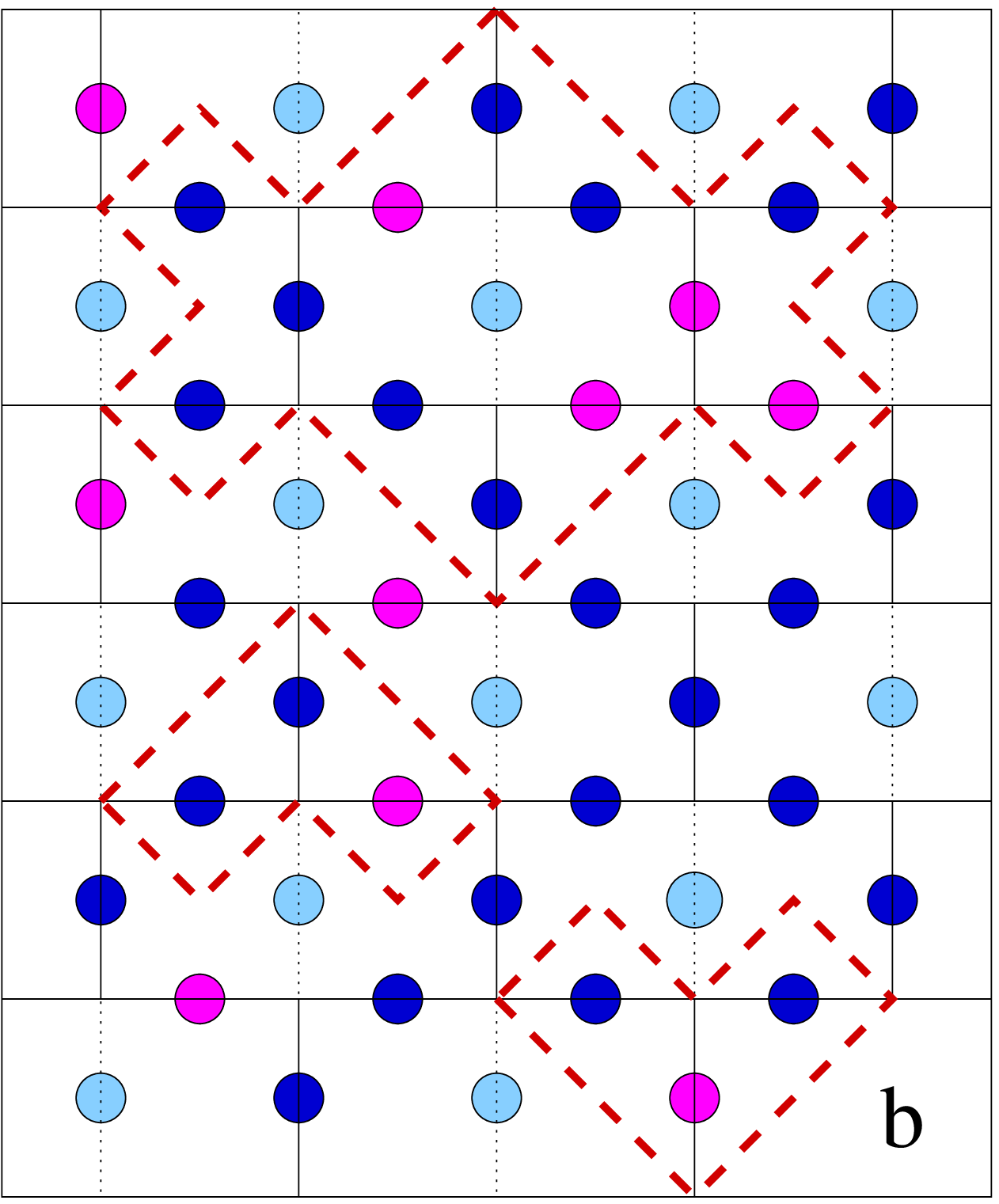}
\includegraphics[width=4.2cm]{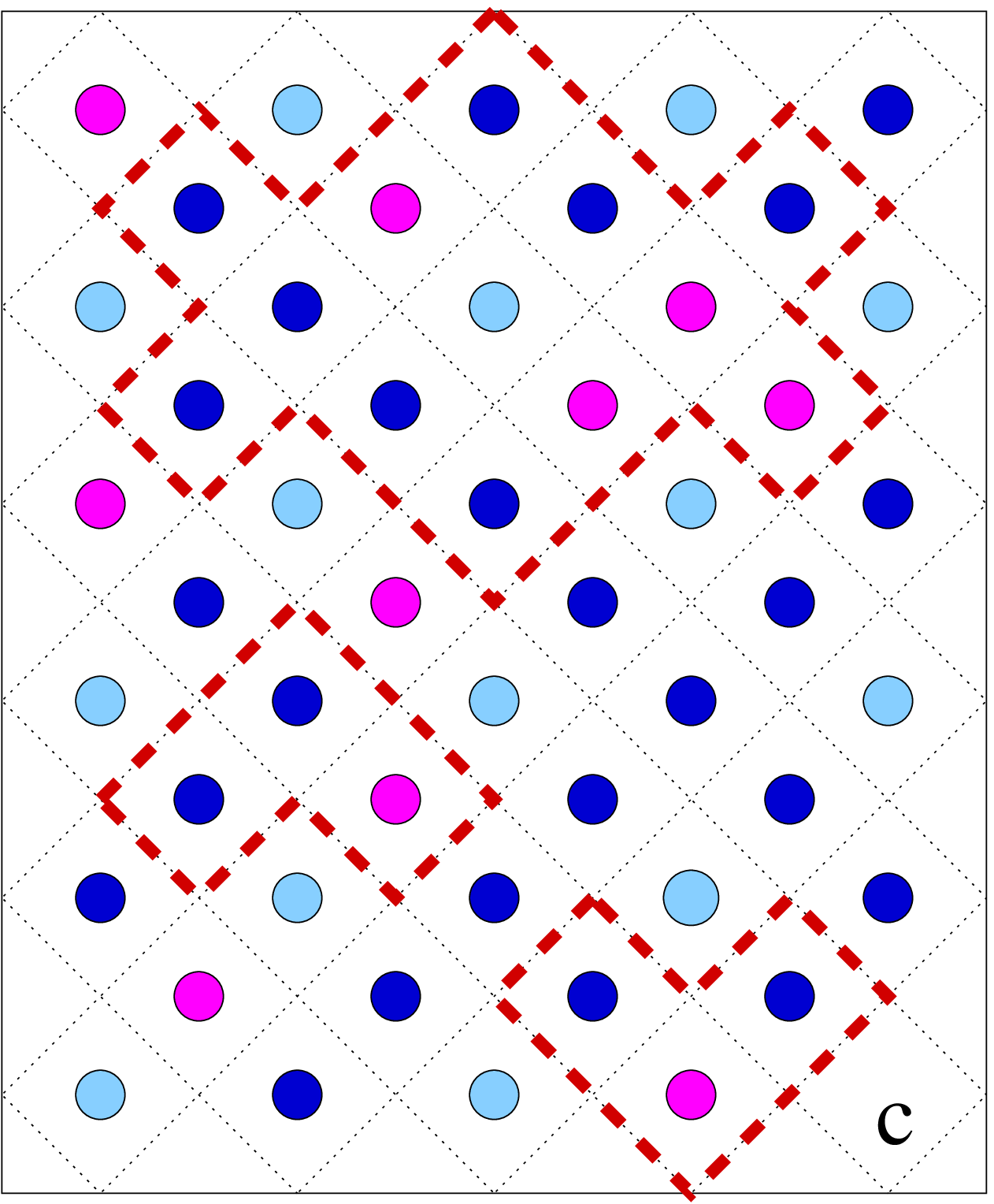} \includegraphics[width=4.2cm]{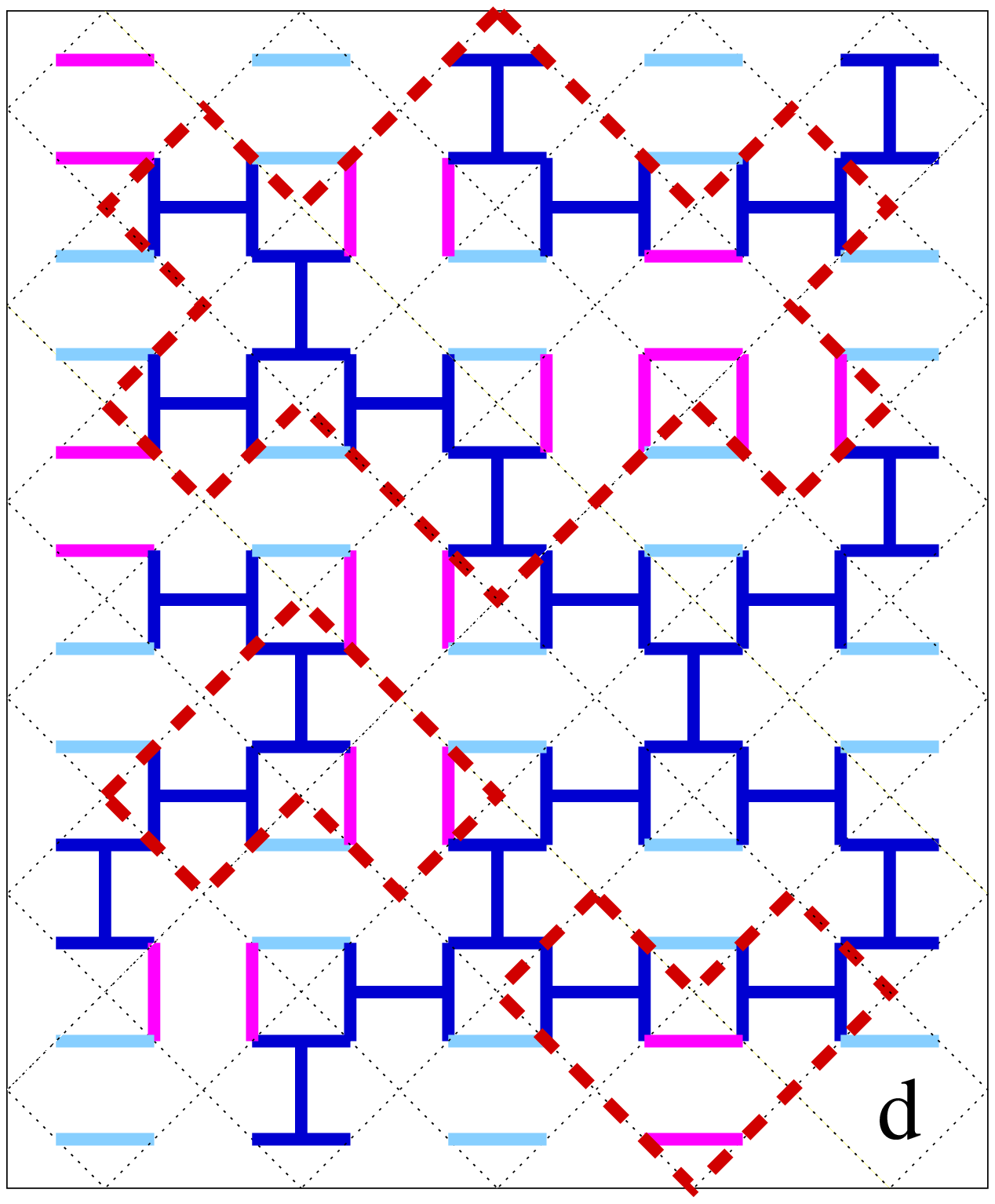}
\includegraphics[width=4.2cm]{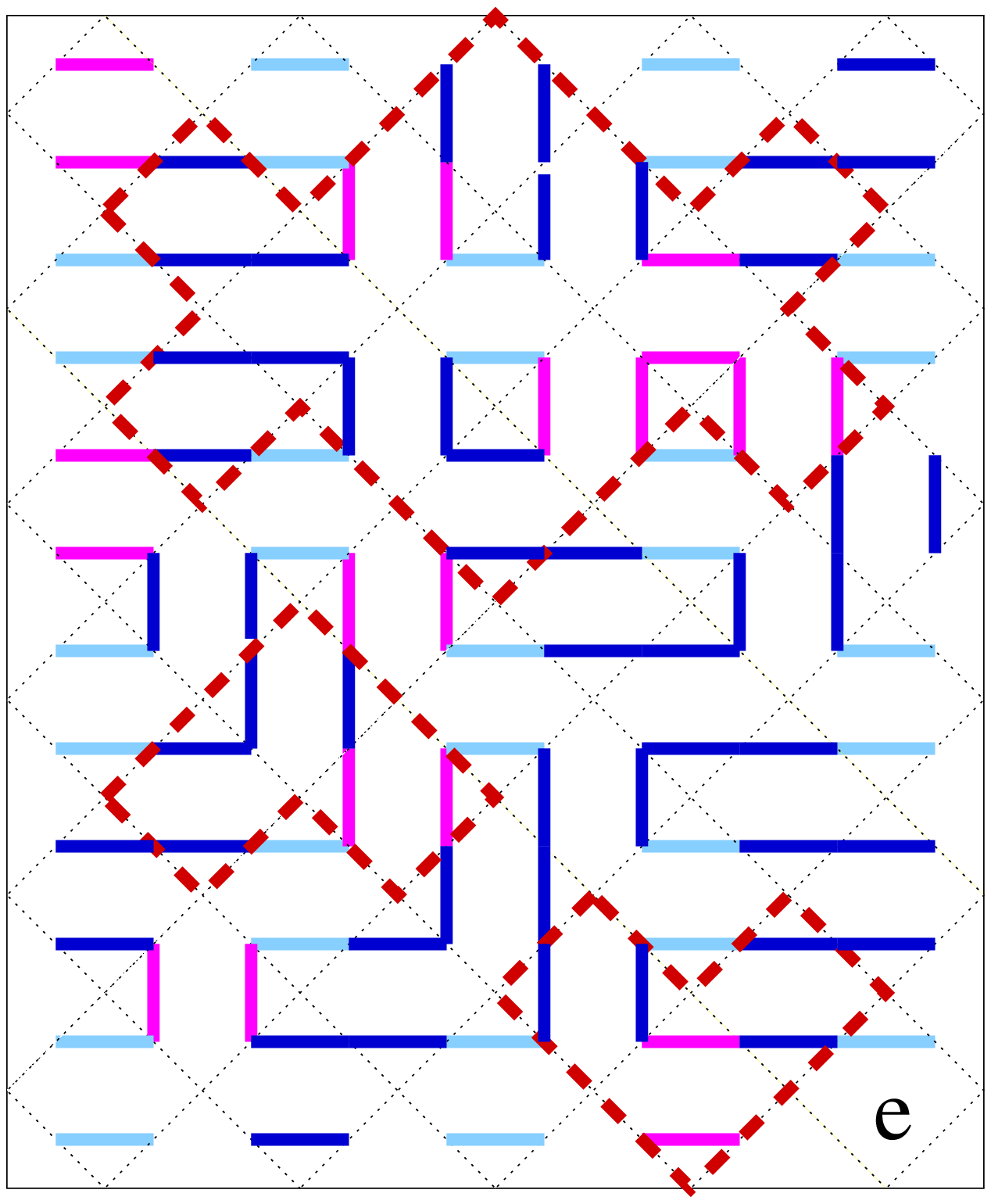} \includegraphics[width=4.2cm]{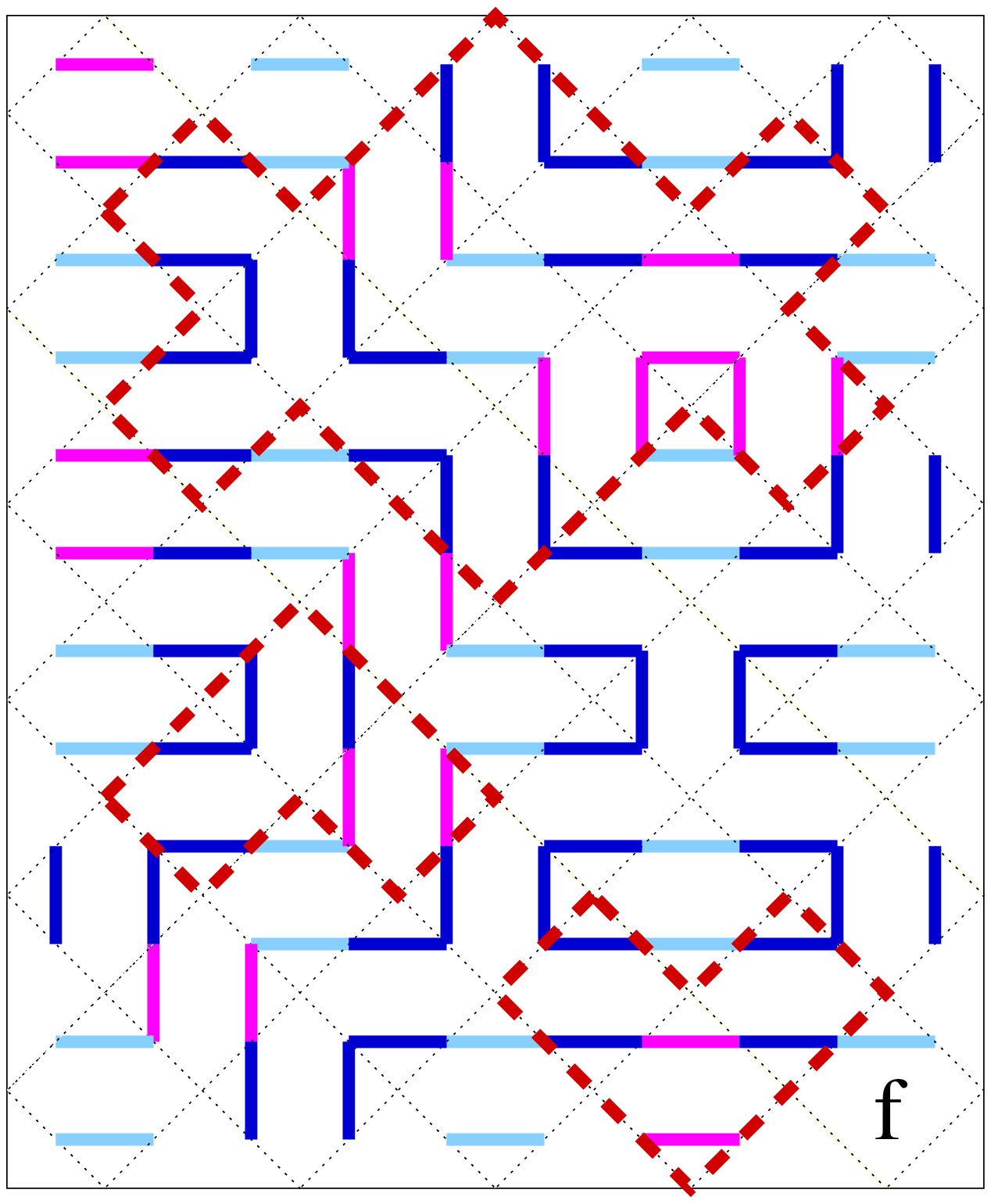}
\caption{(Color online). Exact mapping from  a string-net model on the honeycomb lattice to a fully-packed quantum loop model on the square lattice. Details in the text.}
\end{figure}
Re-drawing the lattice so that the spins are centered on the tiles, we generate the square lattice on which the string-net tile model is defined (FIG.1c). A subtlety worth noticing is that the 12-spin and 3-spin operators can be translated only by an even number of lattice spacings on the square lattice, so that all the operators commute with each other and the dummy spins never take part in the dynamics.

The mapping to a loop model with non-orthogonal inner product proceeds from FIG.1c by choosing local reference systems in a checkerboard fashion, alternating $\{ 
| \hat{\begin{picture}(12,6) \put(2,-2){\includegraphics[width=0.3cm]{conf1.eps}} \end{picture}} \rangle,
| \hat{\begin{picture}(12,6) \put(2,-2){\includegraphics[width=0.3cm]{conf3.eps}} \end{picture}} \rangle \}$ and $\{
| \hat{\begin{picture}(12,6) \put(2,-2){\includegraphics[width=0.3cm]{conf2.eps}} \end{picture}} \rangle,
| \hat{\begin{picture}(12,6) \put(2,-2){\includegraphics[width=0.3cm]{conf4.eps}} \end{picture}} \rangle \}$ as basis vectors. Then, the orthogonal spin-up and spin-down states (\includegraphics[width=0.2cm]{circlem.eps}, \includegraphics[width=0.2cm]{circleb.eps}) can be mapped into the orthogonal states \begin{picture}(12,10) \put(0,-3){\includegraphics[width=0.4cm]{plaq1.eps}} \end{picture} \begin{picture}(12,10) \put(0,-3){\includegraphics[width=0.4cm]{plaq3.eps}} \end{picture} or \begin{picture}(12,10) \put(0,-3){\includegraphics[width=0.4cm]{plaq2.eps}} \end{picture} \begin{picture}(12,10) \put(0,-3){\includegraphics[width=0.4cm]{plaq4.eps}} \end{picture} respectively, outlining on the tiles the topology of the original string-net with dressed vertices (FIG.1d).
By a simple change of basis to the non-orthogonal pair $\{ 
| \hat{\begin{picture}(12,6) \put(2,-2){\includegraphics[width=0.3cm]{conf1.eps}} \end{picture}} \rangle,
| \hat{\begin{picture}(12,6) \put(2,-2){\includegraphics[width=0.3cm]{conf2.eps}} \end{picture}} \rangle \}$, as explained in the first part of the paper, any configuration can finally be written as a linear combination of loop configurations with the correct amplitudes. FIG.1e would be one configuration in such an expansion for the configuration in FIG.1d.

It has been argued \cite{TTSN08} that loop models with orthogonal inner products and local interactions do not support non-Abelian excitations. In FIG.1f we simply re-write the string-net tile model of FIG.1c in terms of loops by choosing local reference systems in a checkerboard fashion, alternating $\begin{picture}(12,10) \put(0,-3){\includegraphics[width=0.4cm]{plaq1.eps}} \end{picture} \; \begin{picture}(12,10) \put(0,-3){\includegraphics[width=0.4cm]{plaq2.eps}} \end{picture}$ and $\begin{picture}(12,10) \put(0,-3){\includegraphics[width=0.4cm]{plaq2.eps}} \end{picture} \; \begin{picture}(12,10) \put(0,-3){\includegraphics[width=0.4cm]{plaq1.eps}} \end{picture}$ as pairs of basis vectors representing the spin-up and spin-down states. Contrarily to what we had so far, here $\begin{picture}(12,10) \put(0,-3){\includegraphics[width=0.4cm]{plaq1.eps}} \end{picture}$ and $ \begin{picture}(12,10) \put(0,-3){\includegraphics[width=0.4cm]{plaq2.eps}} \end{picture}$ are \emph{orthogonal} states and they have \emph{no} relations with the two states $\{ 
| \hat{\begin{picture}(12,6) \put(2,-2){\includegraphics[width=0.3cm]{conf1.eps}} \end{picture}} \rangle,
| \hat{\begin{picture}(12,6) \put(2,-2){\includegraphics[width=0.3cm]{conf2.eps}} \end{picture}} \rangle \}$ and the projector (\ref{HF}). As before, the 12-spin and 3-spin operators can be translated only by an even number of lattice spacings to ensure the exact correspondence with the original string-net model on the honeycomb lattice.
Notice how, geometrically, FIG.1f is nothing but FIG.1a where the strings forming the net are now ribbons and small loops fill the empty spaces as expected from a fully packed model. This last mapping shows explicitly how quantum loop gases with orthogonal inner products \emph{and} local interactions (involving not one, but a finite number of plaquettes) allow for non-Abelian excitations.
In a continuum setting string-nets are then substituted by strongly attracting loops so that each string-net segment is replaced by double stranded loop segments. The dynamics of the loops is by construction identical to the one of string-nets.

\emph{The sign problem and quantum interference.} Although we can now write models of orthogonal loops for non-Abelian anyons, the complexity of the Fibonacci anyons is not overcome, as emphasized in this last section in terms of RK decompositions and sign problems.

Many Hamiltonians admit a Rokhsar-Kivelson (RK) decomposition \cite{RK}. This means they can be written as a weighted sum of projectors $P_{\{c\}}$ connecting the the configurations in the set $\{c\} = \{c_1,\ldots,c_n\}$ for some finite integer $n$:
\begin{displaymath}
H = \sum_{\{c\}} w_{\{c\}} P_{\{c\}}
\end{displaymath}
Being RK-decomposable is a basis-dependent property; however, oftentimes the basis elements are labelled by local degrees of freedom (such as the states of lattice sites, edges or plaquettes) in a natural way; the dynamics can then be implemented in terms of operators connecting the configurations in $\{c\}$ which differ only by one localized degree of freedom. In those cases it is straightforward to write the Hamiltonian in RK form and a unique ground state can be constructed as the state annihilated by all the projectors $P_{\{c\}}$.
If all the projectors have a finite weight $w_{\{c\}}$, then the system is necessarily gapped since any excited state is annihilated by some but not all projectors.

To each Hamiltonian decomposable in terms of $2 \times 2$ projectors one can associate a corresponding classical system \cite{CCMP05,H97,H04,AFF04} and the fact that the ground state is annihilated by projectors implies detailed balance conditions for the transition rates between the corresponding classical configurations.
Kitaev's toric code  \cite{K03} (with its Abelian anyons) provides such an example; even though it admits a different RK decomposition for each of its four degenerate ground states, the degeneracy is detectable only by non-local operators and the ground states are dynamically disconnected in the thermodynamic limit. For the Fibonacci anyons case, instead, each \emph{local} surgery point has a two-fold degenerate ground state (\ref{Falpha}).
Indeed, two non-compatible RK decompositions are possible for the matrix $F_{\alpha}$ in (\ref{Falpha}): one with $2 \times 2$ projectors annihilating $|g_1\rangle$ and the other with projectors annihilating $|g_2\rangle$. As a consequence constructing the ground state wavefunction by following a sequence of states related to each other by simple detailed balance relations is a non-trivial task. This is a simple way to view the complexity of the Fibonacci wavefunction as compared e.g.~to the toric code, whose Hamiltonian can be written as sum of 2-dimensional projectors and related to a classical stochastic system satisfying detailed balance.

A related feature of the $4 \times 4$ Fibonacci projector (\ref{Falpha}) is that not all of its off-diagonal matrix elements have the same sign. Such a ``sign problem'' is the hallmark of quantum-mechanical interference effects as follows:
\begin{displaymath}
\begin{CD}
\begin{picture}(12,-40) \put(-1,-4){\includegraphics[width=0.5cm]{conf1.eps}} \end{picture} @>-\gamma^{-1}>> \begin{picture}(12,-40) \put(-1,-4){\includegraphics[width=0.5cm]{conf2.eps}} \end{picture} \\
@V-\gamma^{-1/2}VV	@VV-\gamma^{-1/2}V \\
\begin{picture}(12,-40) \put(-1,-4){\includegraphics[width=0.5cm]{conf4.eps}} \end{picture} @>\gamma^{-1}>> \begin{picture}(12,-40) \put(-1,-4){\includegraphics[width=0.5cm]{conf3.eps}} \end{picture}
\end{CD}
\qquad \qquad \qquad
\begin{CD}
\begin{picture}(20,-40) \put(-1,-4){\includegraphics[width=0.8cm]{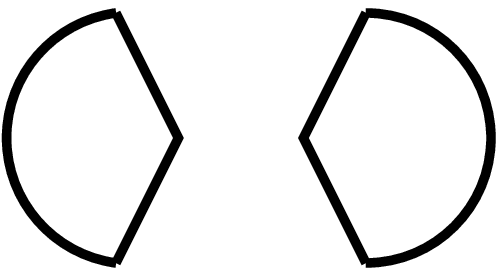}} \end{picture} @>-\gamma^{-1}>> \begin{picture}(20,-40) \put(-1,-4){\includegraphics[width=0.8cm]{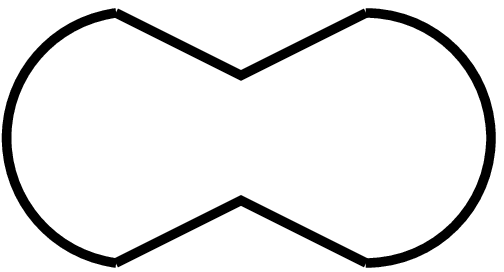}} \end{picture} \\
@V-\gamma^{-1/2}VV	@VV-\gamma^{-1/2}V \\
\begin{picture}(20,16) \put(-2,-3){\includegraphics[width=0.8cm]{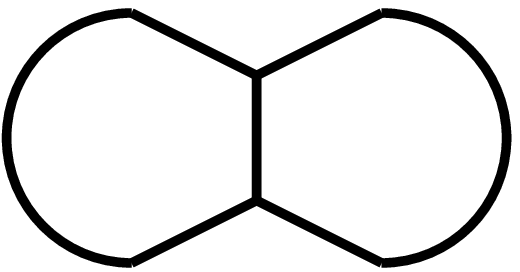}} \end{picture} @>\gamma^{-1}>> \begin{picture}(12,-40) \put(-5,-3){\includegraphics[width=0.8cm]{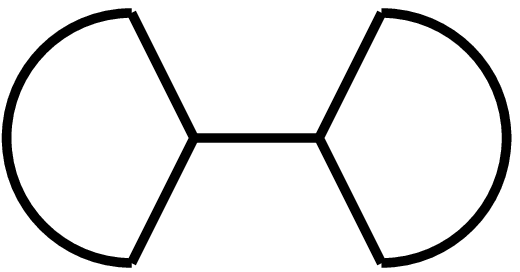}} \end{picture}
\end{CD}
\end{displaymath}
The transition amplitudes above are read off directly from $F$ in (\ref{defF}). In the diagram on the right a choice of boundary conditions has been made: states with and without tadpoles (lower right and upper left corner respectively) are connected by multiple paths which interfere destructively.
The effect of the ``sign problem'' is therefore a \emph{dynamical} partitioning of the Hilbert space into different topological sectors. Note that since $F^n=F$ the interference between \begin{picture}(12,-40) \put(2,-2){\includegraphics[width=0.3cm]{conf1.eps}} \end{picture} and \begin{picture}(12,-40) \put(2,-2){\includegraphics[width=0.3cm]{conf3.eps}} \end{picture} (and between \begin{picture}(12,-40) \put(2,-2){\includegraphics[width=0.3cm]{conf2.eps}} \end{picture} and \begin{picture}(12,-40) \put(2,-2){\includegraphics[width=0.3cm]{conf4.eps}} \end{picture}) is  totally destructive at every order in $F$ and, as a consequence, any argument about the ergodicity of the dynamics cannot rely on purely classical considerations.
At the same time, the sign problem spoils the mapping \cite{CCMP05} of RK-Hamiltonians to classical statistical mechanical systems since a correspondence between the negative off-diagonal elements of the quantum Hamiltonian and the (necessarily positive) classical transition rates cannot be enforced.

\emph{Conclusions.} We have detailed two mappings from string-nets to quantum loop gases with non-orthogonal and orthogonal inner products, showing the equivalence of apparently diverse models. The non-orthogonality of quantum loop states arises in a transparent way when hard topological constraints are present and the description is restricted to a Hilbert space of reduced dimensionality. Non-orthogonal inner products for quantum loops gases are, however, not necessary to generate non-Abelian anyons and orthogonal inner products can be used as well if the appropriate dynamics is defined. Whether or not a non-Abelian Fibonacci phase can be described with interactions involving fewer than 12 spins remains an open problem.

\emph{Acknowledgments.}
We are grateful to Paul Fendley and Kirill Shtengel for useful discussions. This work was supported by DOE Grant DEFG02-06ER46316 (A. V. and C. C.) and NSF Grant DMR-0706078 (X.-G. W.).

\end{document}